\ifdefined\WORDCOUNT
\documentclass[superscriptaddress,twocolumn,preprintnumbers,amsmath,amssymb,prl,nofootinbib]{revtex4-1}
\else
\documentclass[superscriptaddress,twocolumn,showpacs,preprintnumbers,amsmath,amssymb,prl]{revtex4-1}
\fi

\usepackage{amssymb}
\usepackage{epsfig,amsmath,graphics}
\usepackage{epstopdf}
\usepackage{graphicx}

\newcommand{\ket}[1]{\ensuremath{\left|#1\right>}}

\ifdefined\WORDCOUNT
    \newcommand{\HideEq}[1]{}
\else
    \newcommand{\HideEq}[1]{#1}
\fi

\def\be{\begin{equation}}
\def\ee{\end{equation}}

\begin{document}

\title{Phase shift in atom interferometry due to spacetime curvature}

\def\stanfordAffiliation{Department of Physics, Stanford University, Stanford, California 94305}

\author{Peter Asenbaum}
\affiliation{\stanfordAffiliation}
\author{Chris Overstreet}
\affiliation{\stanfordAffiliation}
\author{Tim Kovachy}
\affiliation{\stanfordAffiliation}
\author{Daniel D. Brown}
\affiliation{School of Physics and Astronomy, University of Birmingham, Birmingham, B15 2TT, UK}
\author{Jason M. Hogan}
\affiliation{\stanfordAffiliation}
\author{Mark A. Kasevich}
\affiliation{\stanfordAffiliation}
\email{kasevich@stanford.edu}

\date{\today}

\begin{abstract}
We present a single-source dual atom interferometer and utilize it as a gradiometer for precise gravitational measurements.  The macroscopic separation between interfering atomic wave packets (as large as 16\,cm) reveals the interplay of recoil effects and gravitational curvature from a nearby Pb source mass.  The gradiometer baseline is set by the laser wavelength and pulse timings, which can be measured to high precision. Using a long drift time and large momentum transfer atom optics, the gradiometer reaches a resolution of $3 \times 10^{-9}$\,s$^{-2}$ per shot and measures a 1\,rad phase shift induced by the source mass.
\end{abstract}

\pacs{03.75.-b, 37.25.+k, 91.10.Pp, 03.65.Ta}

\ifdefined\WORDCOUNT
  
\else
   \maketitle
\fi

Light-pulse atom interferometers are ideally suited to study the interaction between gravity and quantum particles.  Previous atomic gravimeters \cite{Kasevich1992,Peters1999} and gravity gradiometers \cite{McGuirk2002,Sorrentino2012a,Biedermann2015} may be understood through a classical analogy in which the lasers act as rulers that measure atomic positions.  In quantum mechanics, such position measurements inevitably impart momentum to the atoms.
This momentum transfer perturbs the atomic trajectories, causing the atoms to experience a different gravitational force if gravitational curvature is present.  A longstanding goal in matter-wave interferometry has been to resolve the phase shifts induced by this coupling of recoil effects to gravitational curvature \cite{Anandan1984,Audretsch1994,Borde1996,Marzlin1996}.

In classical physics, it is always possible to define the local Lorentz frame \cite{Carroll2004a} in which the spacetime around a particle is locally flat.  In quantum mechanics, however, a particle may be placed in a superposition state in which its wave function is delocalized over a macroscopic distance.  In the presence of gravitational curvature, there is no coordinate system in which the entire wave function experiences locally flat spacetime.  The two arms of an interferometer operated in this regime will experience different local gravitational forces, which can lead to a phase shift that characterizes the gravitational curvature \footnote{Previous work that measured the curvature of the gravitational field (rather than potential, as in this work) using three separate gravimeters \cite{Rosi2015} was not based on recoil effects}.

The sensitivity of an atom interferometer to acceleration, which scales with its enclosed spacetime area, can be improved by increasing the interferometer time $T$ \cite{Dickerson2013, Muntinga2013} or by utilizing large momentum transfer (LMT) atom optics \cite{McGuirk2000,Muller2008,Chiow2011}.  These techniques have been demonstrated simultaneously \cite{Kovachy2015a}, and we now use them to perform a high-resolution measurement.  Long-T LMT interferometers have been anticipated to reveal the influence of gravitational curvature on the evolution of quantum states \cite{Anandan1984,Audretsch1994,Borde1996,Marzlin1996}.  In addition, the macroscopic wave packet separation in long-T LMT interferometers enables tests of quantum mechanics at large scales \cite{Arndt2014,Nimmrichter2013}.

In this Letter, we demonstrate a long-T LMT dual interferometer generated from a single atomic source.  Because they are highly sensitive to inertial effects, long-T LMT interferometers are especially susceptible to phase noise arising from vibrations in the laser delivery optics \footnote{In the apparatus described here, vibration of the retro-reflection mirror used for the atom optics laser beams is the primary source of phase noise \cite{Dickerson2013}}.  A dual interferometer allows the vibration-induced phase noise to be cancelled as a common mode, with the first interferometer acting as a phase reference for the second \cite{McGuirk2002}.  In combination with the large intrinsic sensitivity of long-T LMT interferometers, this phase noise rejection makes the dual interferometer presented here an ideal platform for a wide range of measurements, such as the gravity gradiometry results discussed in this work.  Additionally, the dual interferometer appears to be well suited for precision measurements of atomic polarizabilities \cite{Gregoire2015} and tune-out wavelengths \cite{Holmgren2012,Leonard2015} and tests of atom charge neutrality \cite{Arvanitaki2008}.  In these applications, a key advantage is that the large separation between interfering wave packets would allow individual interferometer arms to be independently addressed for long times.

Previous atomic gravity gradiometers \cite{McGuirk2002,Sorrentino2012a,Biedermann2015}, which used independently generated atom clouds separated by a baseline as the sources for two accelerometers, were subject to uncertainty in the baseline length due to source position fluctuations.  In contrast, the baseline of the gradiometer presented here is insensitive to the atom source position.  The gradiometer operates in a macroscopic regime in which the interferometer arms experience resolvably different forces and confirms the importance of recoil effects in precision inertial sensing.  Additionally, our observation of a stable differential phase between two macroscopic matter wave interferometers separated over a long baseline constrains phase noise due to exotic extensions of quantum mechanics.

\begin{figure}
\begin{center}
\includegraphics[width=\columnwidth]{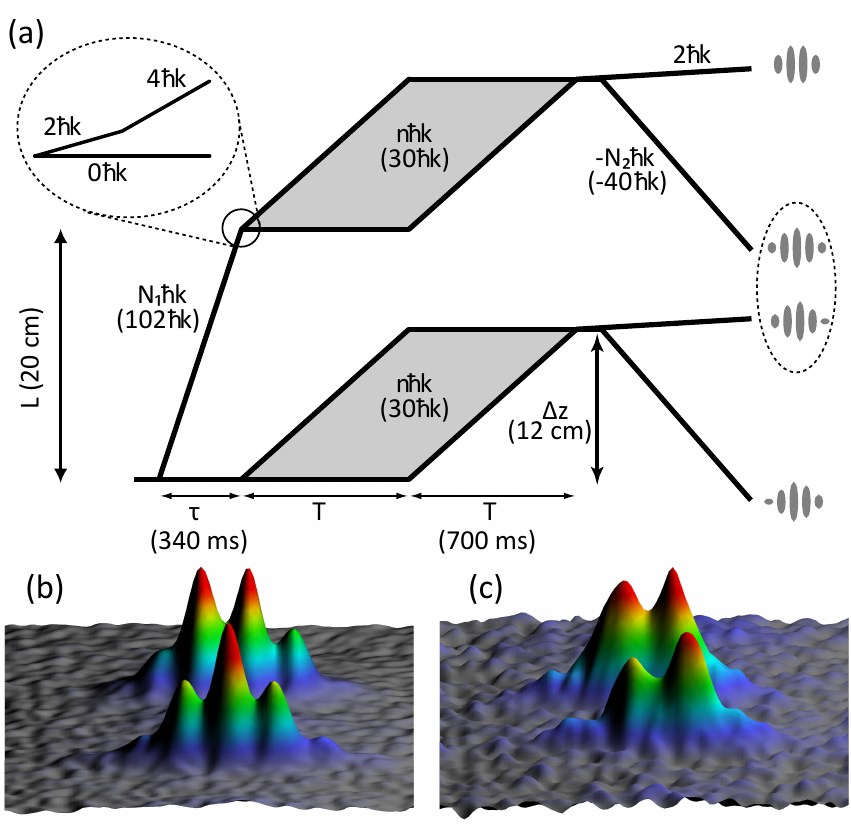}
\caption{ \label{Fig:Schematic} (a)  Spacetime diagram of the centers of the wave packets for a gravity gradiometer sequence, neglecting gravitational acceleration.  An initial atom source is split into two clouds that drift apart with momentum difference $N_1 \hbar k$ for time $\tau$, ultimately separating by the baseline $L$.  Each cloud is used as the source for a Mach-Zehnder interferometer with momentum splitting $n \hbar k$ and pulse spacing $T$.  The path separation between the two interferometer arms reaches a distance of $\Delta z = (n \hbar k/m) T$.  The numerical values in parentheses indicate typical experimental parameters.  In this work, $k$ is negative for the initial splitting and interferometer pulses, i.e. the interferometer trajectories are below the unperturbed launch height.  Data plots:  fluorescence images of spatial interference fringes for an interferometer with (b) $\Delta z = 4$\,cm ($n=10$) and (c) $\Delta z = 12$\,cm ($n=30$).}
\end{center}
\end{figure}

Each experimental run begins with the preparation of an ultracold $^{87}$Rb atom cloud \cite{Kovachy2015a}.
Following evaporative cooling, the velocity spread of the atom ensemble is further reduced by a magnetic lensing sequence \cite{Kovachy2015}.  The atoms are then transferred into the magnetically insensitive $\ket{F=2, m_{\text{F}} = 0}$ state by a microwave pulse and launched vertically into a 10\,m tall atomic fountain using an optical lattice.  Subsequently, the atoms undergo a secondary collimating lensing sequence in the transverse dimensions, with the lensing force provided by the transverse intensity profile of a 1\,mm waist, red-detuned laser beam (optical dipole lens).  This second lens is applied $\sim$200\,ms after the end of the magnetic lens.  Ultimately, the launched atom cloud contains $\sim$$10^{6}$ atoms with an effective temperature of $\sim$\,50\,nK in the transverse dimensions.

The LMT atom optics used in this work consist of sequences of two-photon Bragg transitions \cite{Kovachy2015a}.  Figure \ref{Fig:Schematic}(a) illustrates the spacetime diagram associated with a single-source gradiometer sequence.  An initial LMT beam splitter sequence splits the atom cloud into two wave packets with momenta differing by $N_1$ photon momentum recoil kicks ($N_1 \hbar k$, where $k$ is the wave number of the laser used to drive the Bragg transitions) in the vertical direction.  This beam splitter sequence consists of an initial $\pi/2$-pulse to generate two wave packets split by momentum $2 \hbar k$ followed by $(N_1/2 - 1)$ sequential $\pi$-pulses to further accelerate the wave packet that receives the initial momentum kick \cite{Kovachy2015a}.  The wave packets are allowed to freely drift apart for a time $\tau$.  Next, the initially accelerated arm is decelerated by a sequence of $(N_1/2 - 1)$ $\pi$-pulses, so that the momentum splitting between the wave packets is reduced to $2 \hbar k$.  The two wave packets are vertically separated by a baseline $L = (N_1 \hbar k/m) \tau$ (where $m$ is the atomic mass) and are the respective sources for the dual interferometers used in the gradiometer \footnote{The finite duration of the optical pulses leads to small corrections to calculated distances.}.  Before the initial beam splitter sequence, the vertical velocity distribution is filtered by two long-duration $\pi$-pulses (Gaussian temporal profile, FWHM 200\,$\mu$s).

The interferometers are initiated by a beam splitter sequence like the one used to split the initial atom cloud (the two vertically displaced wave packets use opposite input ports of the first interferometer beam splitter, since their momenta differ by $2 \hbar k$).  We use a Mach-Zehnder interferometer sequence with pulse spacing $T$.  The momentum difference between the interferometer arms is denoted by $n \hbar k$.  Midway through the interferometer, a sequence of $(n-1)$ $\pi$-pulses exchanges the momenta of the two arms, effectively acting as a mirror that redirects the two arms back toward each other.  The final beam splitter sequence consists of $(n/2 - 1)$ $\pi$-pulses followed by a $\pi/2$-pulse to overlap the arms in momentum.  The laser system and optics configuration used to drive the Bragg transitions is described in \cite{Kovachy2015a}.

To measure the differential phase shift between the two interferometers (gradiometer phase), we image one output port from each interferometer onto a CCD camera using resonant scattering.  Because of the large vertical displacement between the two interferometers, we deliver an additional momentum kick $-N_2 \hbar k$ to the lower port of each interferometer, so that the lower port of the upper interferometer and the upper port of the lower interferometer fit into the CCD camera's field of view at the time of detection [see Fig. \ref{Fig:Schematic}(a)].  We use phase shear readout \cite{Sugarbaker2013,Muntinga2013} to extract a value for the gradiometer phase from each individual run of the experiment.  Specifically, the angle of the Bragg laser beams is slightly tilted for the final beamsplitter sequence using a piezo tip-tilt stage on the retro-reflection mirror, imprinting a horizontal phase gradient across the cloud.  This leads to horizontal spatial fringes in the interferometer output ports [see Fig. \ref{Fig:Schematic}(b)-(c)], allowing for single-shot determination of phase and contrast \cite{Sugarbaker2013} in a single port. The gravity gradient is then
calculated from the gradiometer phase $d\phi$ using $T_{zz} \approx -d\phi/(n \hbar k L T^2)$. See \cite{supplemental} for a description of the data analysis process.

The gradiometer achieves a resolution of 3\;E per shot with parameters $L=32$\,cm, $n=20$, and $T=600$\,ms ($1\,\text{E} = 10^{-9}\;\text{s}^{-2}$) \cite{supplemental}.
This is near the estimated shot noise limit of $\sim 1$\,E per shot.   Improvements in the atom source and imaging system would increase the atom number and contrast, allowing better resolution.
We implement gradient measurements using interferometers with path separations of up to $\Delta z = 16$\,cm with $L=20$\,cm, $n=38$, and $T=700$\,ms.

\begin{figure}
\begin{center}
\includegraphics[width=\columnwidth]{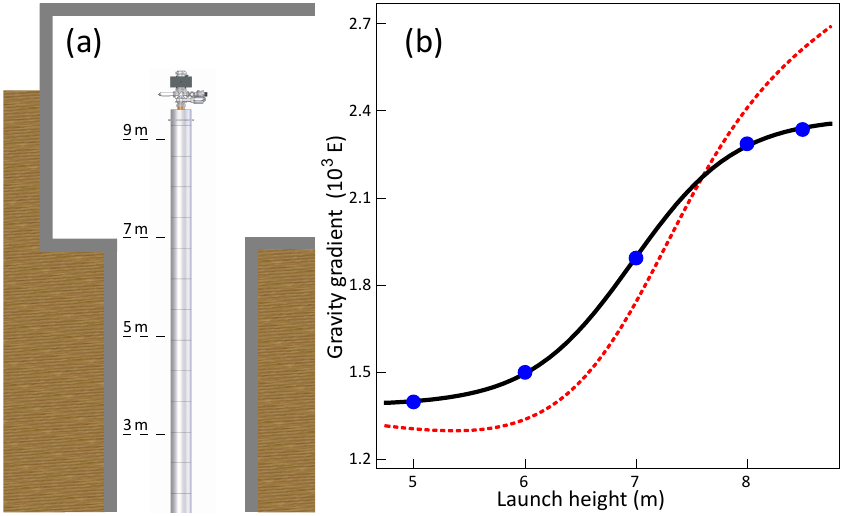}
\caption{ \label{Fig:GGvsLaunchHeight} (a) Schematic of the interferometer and surrounding mass, to scale, with indicated launch heights.  (b) Measured gravity gradient (blue points) as a function of launch height $h$ for a gradiometer with $L = 10$\,cm, $n = 10$, and $T = 500$\,ms.  The red, dashed curve is a model of the interferometer response to the gravity gradient produced by a spherical Earth, including the effects of the cylindrical pit and basement in which the interferometer is located.  The model has one free parameter, the density of Earth's surface near the interferometer, which fits to $\rho = 2.3$\,g/cm$^3$.  The black, solid curve is an empirically-motivated logistic fit.  Statistical uncertainties are $\approx$\hspace{0.02in}6 E (error bars smaller than data points).}
\end{center}
\end{figure}

\begin{figure}
\begin{center}
\includegraphics[width=\columnwidth]{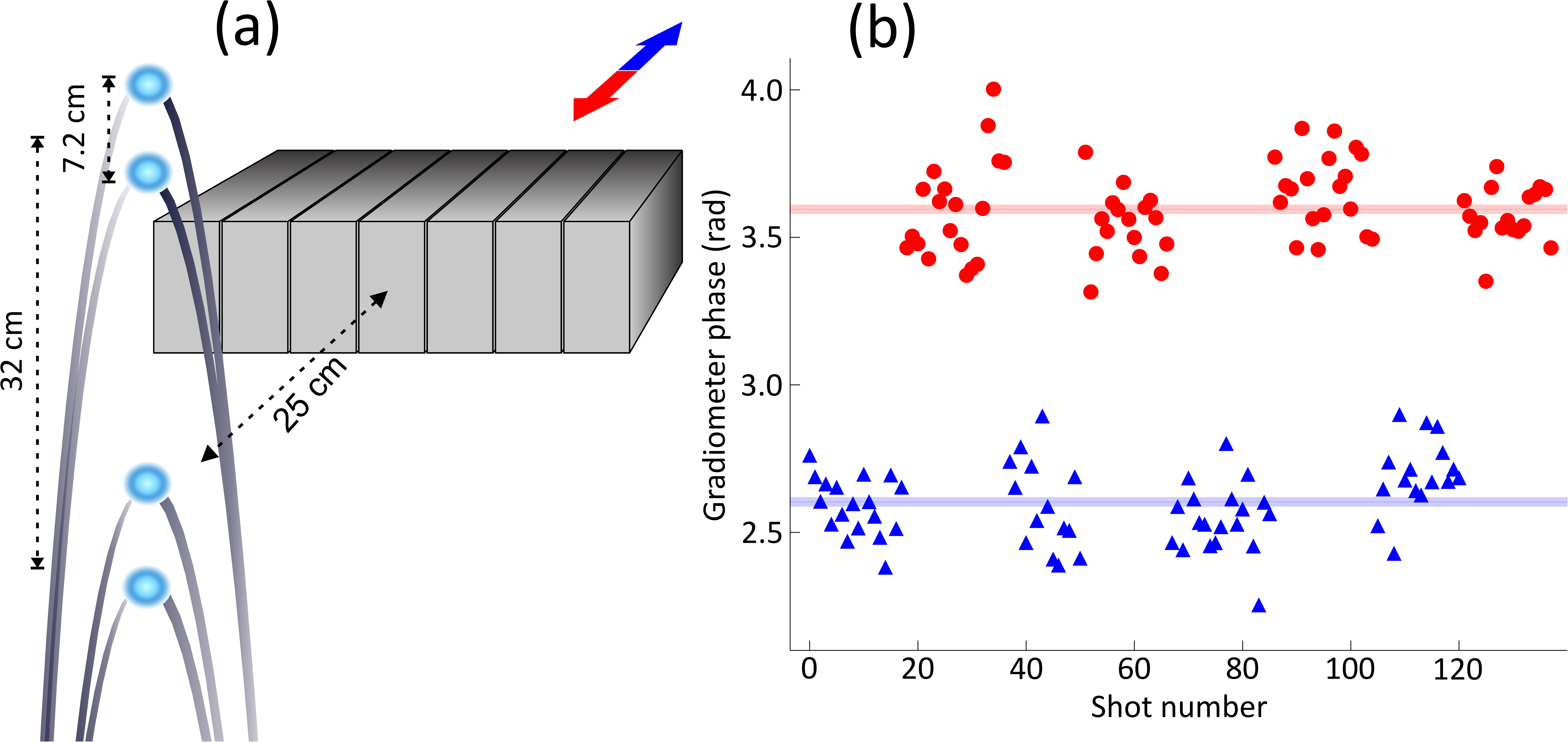}
\caption{ \label{Fig:BricksPhasevLaunchHeight} (a)  Schematic representation of the experimental setup for measuring the gravity gradient of seven lead bricks (total mass $84$\,kg).  (b)  Measured gradiometer phase of a sequence with $h = 8.45$\,m, $L = 32$\,cm, $n = 20$, and $T = 600$\,ms, with (red) and without (blue) the bricks present.  }
\end{center}
\end{figure}

\begin{figure}
\begin{center}
\includegraphics[width=\columnwidth]{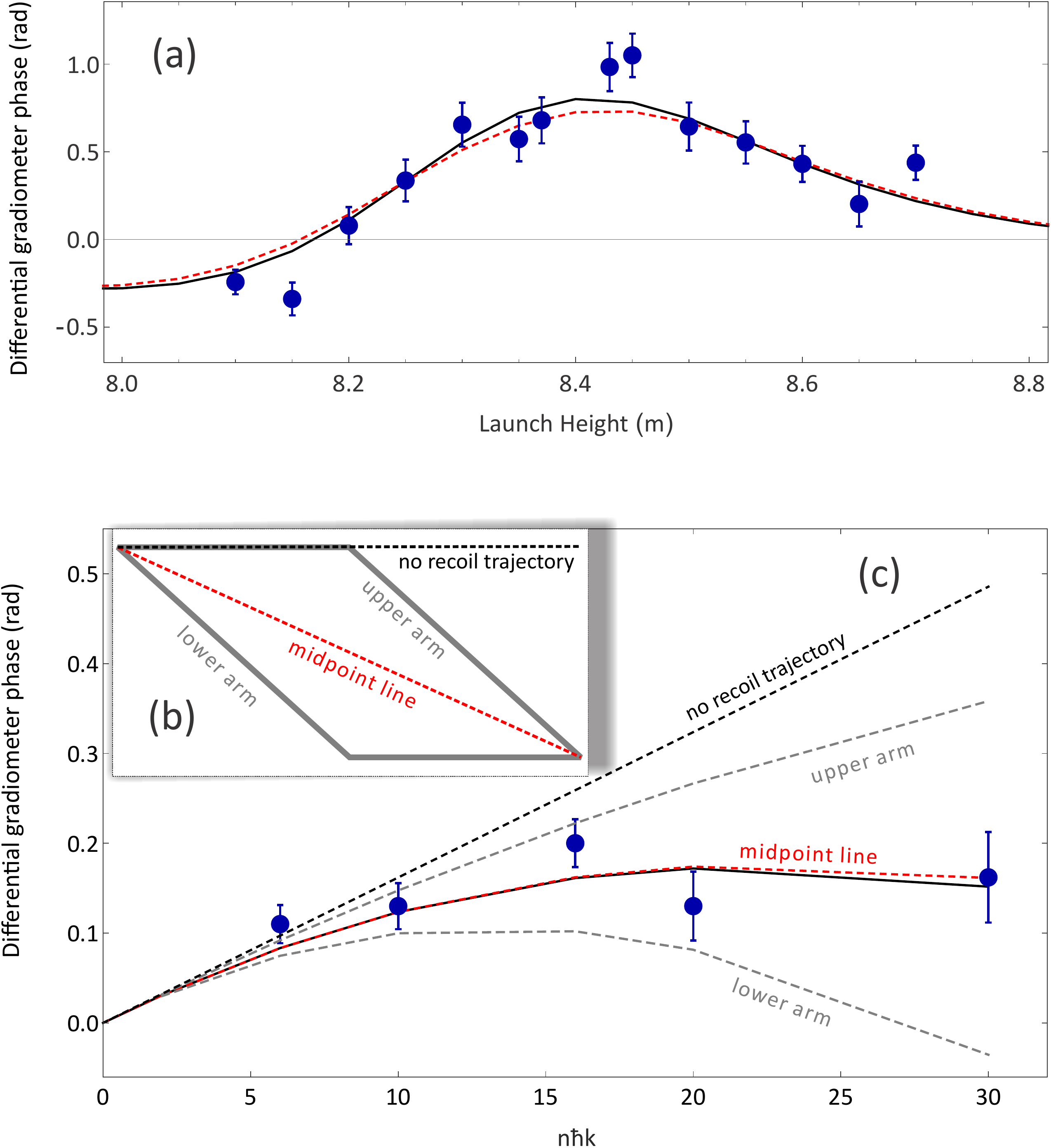}
\caption{ \label{Fig:BricksPhasevLaunchHeight2} (a) Differential gradiometer phase (with and without bricks present) as a function of launch height with $L = 10$\,cm, $n = 30$, and $T = 900$\,ms ($\Delta z = 16$\,cm).  The red curve is the midpoint line prediction and the black curve the full phase shift calculation. (b) Schematic of the interferometer trajectories, neglecting gravitational acceleration:  upper and lower arm (gray), midpoint line (red) and the trajectory without recoil effects (dashed black). (c) Differential gradiometer phase of a sequence with $h = 8.25$\,m, $L = 32$\,cm, and $T = 550$\,ms as a function of $n$ ($\Delta z = 2$\,cm to $\Delta z = 10$\,cm).  The black, dashed curve is the phase shift calculated without including recoil effects.  The upper and lower dashed curves are the phase shifts calculated by using the gravitational forces on the upper and lower interferometer arms, respectively. The red (solid black) curve is the phase predicted by the midpoint line (full phase shift calculation).}
\end{center}
\end{figure}

The gravity gradient can be measured as a function of vertical position by varying the lattice launch velocity.  Fig.~\ref{Fig:GGvsLaunchHeight} shows the measured gravity gradient as a function of launch height.  The observed spatial variation of the gravity gradient is reasonably consistent with a model that includes the PREM (preliminary reference earth model) \cite{Dziewonski1981}, the cylindrical pit surrounding the interferometer, and the basement in which the lab is located. An empirical fit to the measured gravity gradient is used to predict the gradiometer phase as a function of launch height $h$ for other measurements, such as those described by Figs.~\ref{Fig:BricksPhasevLaunchHeight} and~\ref{Fig:BricksPhasevLaunchHeight2}.

To illustrate the gradiometer's sensitivity to nearby mass, we placed several lead bricks near the apex of the interferometer trajectory and observed their effect on the gradiometer phase  (Fig.~\ref{Fig:BricksPhasevLaunchHeight}).  The bricks produce a phase shift of 1.0\,rad, in agreement with the theoretical prediction obtained by numerically calculating the propagation, laser, and separation phases along the perturbed interferometer trajectories \cite{Hogan2009} (Fig.~\ref{Fig:BricksPhasevLaunchHeight}). This sensitivity makes the experimental setup a promising candidate for measuring the gravitational constant G \cite{Fixler2007a, Rosi2014}. We find the systematic error of the gradiometer phase due to changes in launch position to be small  \cite{supplemental}.

Figure ~\ref{Fig:BricksPhasevLaunchHeight2}(a) compares the difference in the gradiometer phase (with and without bricks present) to its predicted value as a function of launch height.  We note that Bord\'e's midpoint theorem \cite{Antoine2003}, which is valid for gravitational potentials of degree 2 and below, provides a good approximation (red curve) of the full gradiometer phase shift (black curve).

The macroscopic spatial and temporal scales of our interferometers allow the interferometers to resolve phase shifts from the coupling of atomic recoil effects to the curvature of the gravitational field.   The momentum recoil kicks that the atoms receive during the beam splitter and mirror interactions lead to spatial trajectory perturbations with characteristic size equal to the interferometer path separation $\Delta z = (n \hbar k/m) T$.  In the presence of a nonuniform gravitational field, these trajectory perturbations lead to additional phase shifts.  For a single interferometer in a uniform gravity gradient $T_{zz}$, the recoil phase shift is $\Delta \phi_\text{R} = -(\hbar/2m) n^2 k^2 T_{zz} T^3 = -(1/2) n k \Delta z T_{zz} T^2$ \cite{Hogan2009}.  Since $\Delta \phi_\text{R}$ grows rapidly with $n$ and with $T$, our interferometer parameters are well-suited to resolve it.

For a uniform gravity gradient, $\Delta \phi_\text{R}$ would be the same for the upper and lower interferometers and thus would not be present in the differential phase shift.  To circumvent this limitation, we use the gravitational field of the lead bricks, which has a spatially varying gradient.  Figure~\ref{Fig:BricksPhasevLaunchHeight2}(c) shows the difference between the gradiometer phase measured with and without bricks as a function of $n$ .  These measurements are compared to several theory curves: the phase shift calculated ignoring recoil effects, the phase shift calculated using the gravitational forces along the upper (or lower) interferometer arms, and the actual phase shift including all contributions.  To ignore recoil effects in the calculation, we artificially set the atomic mass to be infinite.  The measurements deviate strongly from this theory curve, demonstrating that recoil effects are an important contribution to the phase shift.  Without recoil effects, the phase shift would increase linearly with $n$.  To calculate the phase shift using the gravitational forces along the upper (or lower) interferometer arms, we take the phase shift for each interferometer to be equal to what an interferometer with infinitely massive particles would measure if those particles were subject to the local gravitational accelerations along the upper (or lower) arms of our actual interferometers.  The resulting theory curve using the upper arms is resolvably different from the theory curve using the lower arms, indicating that that the upper and lower interferometer arms experience resolvably different forces.

The data are consistent with the theory curve using the full phase shift calculation.  For our experimental parameters, the same theory curve can also be reproduced (to a good approximation) by assuming interferometers with infinitely massive particles that travel along the midpoint line. The correctly calculated phase shift is also approximately the average of the theory curve using the force along the upper arms and the theory curve using the force along the lower arms.  One can therefore think of the recoil phase shift in our experiment as arising from the deviation in the average gravitational force experienced by the atoms due to the trajectory deflections from the atom optics interactions.  It is worth noting that no atoms actually travel along the interferometer midpoint line. Therefore, the measured phase shift is not determined by the force along one populated trajectory, giving the recoil phase shift a nonlocal character.

The observed phase stability of our gradiometer (e.g., 130\,mrad per shot for $L = 32$\,cm, $n=30$, and $T = 550$\,ms) can be used to constrain extensions of quantum mechanics that would manifest themselves through additional noise in the gradient measurement, due to, for example, anomalous wave packet localization at a length scale of $\sim$10\;cm \cite{Nimmrichter2013}.  Bounding the phase noise of widely-separated, macroscopic interferometers is complementary to previous work \cite{Kovachy2015a,Stamper-Kurn2016,Kovachy2016}, which was designed to be sensitive to spurious phase shifts that would occur inhomogeneously  \cite{Kovachy2015a,Kovachy2016}.  With a suitable source mass configuration, this gravity gradiometer could be used to measure the gravitational Aharonov-Bohm effect \cite{Hohensee2012} and the gravitational constant \cite{Fixler2007a, Rosi2014}.

\ifdefined\WORDCOUNT
\else
\begin{acknowledgments}
We thank M Squared Lasers for loaning us a SolsTiS laser.  C.O. acknowledges support from the Stanford Graduate Fellowship. This work was supported in part by NASA GSFC Grant No. NNX11AM31A.  D.D.B. would like to thank the EC project Q-Sense (project number 691156) within the H2020 framework Quantum Hub for Sensors and Metrology (EPSRC funding within grant EP/M013294/1).
\end{acknowledgments}
\fi


\begin{thebibliography}{37}%
\makeatletter
\providecommand \@ifxundefined [1]{%
 \@ifx{#1\undefined}
}%
\providecommand \@ifnum [1]{%
 \ifnum #1\expandafter \@firstoftwo
 \else \expandafter \@secondoftwo
 \fi
}%
\providecommand \@ifx [1]{%
 \ifx #1\expandafter \@firstoftwo
 \else \expandafter \@secondoftwo
 \fi
}%
\providecommand \natexlab [1]{#1}%
\providecommand \enquote  [1]{``#1''}%
\providecommand \bibnamefont  [1]{#1}%
\providecommand \bibfnamefont [1]{#1}%
\providecommand \citenamefont [1]{#1}%
\providecommand \href@noop [0]{\@secondoftwo}%
\providecommand \href [0]{\begingroup \@sanitize@url \@href}%
\providecommand \@href[1]{\@@startlink{#1}\@@href}%
\providecommand \@@href[1]{\endgroup#1\@@endlink}%
\providecommand \@sanitize@url [0]{\catcode `\\12\catcode `\$12\catcode
  `\&12\catcode `\#12\catcode `\^12\catcode `\_12\catcode `\%12\relax}%
\providecommand \@@startlink[1]{}%
\providecommand \@@endlink[0]{}%
\providecommand \url  [0]{\begingroup\@sanitize@url \@url }%
\providecommand \@url [1]{\endgroup\@href {#1}{\urlprefix }}%
\providecommand \urlprefix  [0]{URL }%
\providecommand \Eprint [0]{\href }%
\providecommand \doibase [0]{http://dx.doi.org/}%
\providecommand \selectlanguage [0]{\@gobble}%
\providecommand \bibinfo  [0]{\@secondoftwo}%
\providecommand \bibfield  [0]{\@secondoftwo}%
\providecommand \translation [1]{[#1]}%
\providecommand \BibitemOpen [0]{}%
\providecommand \bibitemStop [0]{}%
\providecommand \bibitemNoStop [0]{.\EOS\space}%
\providecommand \EOS [0]{\spacefactor3000\relax}%
\providecommand \BibitemShut  [1]{\csname bibitem#1\endcsname}%
\let\auto@bib@innerbib\@empty
\bibitem [{\citenamefont {Kasevich}\ and\ \citenamefont
  {Chu}(1992)}]{Kasevich1992}%
  \BibitemOpen
  \bibfield  {author} {\bibinfo {author} {\bibfnamefont {M.}~\bibnamefont
  {Kasevich}}\ and\ \bibinfo {author} {\bibfnamefont {S.}~\bibnamefont {Chu}},\
  }\href@noop {} {\bibfield  {journal} {\bibinfo  {journal} {Appl. Phys. B
  Photophysics Laser Chem.}\ }\textbf {\bibinfo {volume} {54}},\ \bibinfo
  {pages} {321} (\bibinfo {year} {1992})}\BibitemShut {NoStop}%
\bibitem [{\citenamefont {Peters}\ \emph {et~al.}(1999)\citenamefont {Peters},
  \citenamefont {Chung},\ and\ \citenamefont {Chu}}]{Peters1999}%
  \BibitemOpen
  \bibfield  {author} {\bibinfo {author} {\bibfnamefont {A.}~\bibnamefont
  {Peters}}, \bibinfo {author} {\bibfnamefont {K.~Y.}\ \bibnamefont {Chung}}, \
  and\ \bibinfo {author} {\bibfnamefont {S.}~\bibnamefont {Chu}},\ }\href@noop
  {} {\bibfield  {journal} {\bibinfo  {journal} {Nature}\ }\textbf {\bibinfo
  {volume} {400}},\ \bibinfo {pages} {849} (\bibinfo {year}
  {1999})}\BibitemShut {NoStop}%
\bibitem [{\citenamefont {McGuirk}\ \emph {et~al.}(2002)\citenamefont
  {McGuirk}, \citenamefont {Foster}, \citenamefont {Fixler}, \citenamefont
  {Snadden},\ and\ \citenamefont {Kasevich}}]{McGuirk2002}%
  \BibitemOpen
  \bibfield  {author} {\bibinfo {author} {\bibfnamefont {J.~M.}\ \bibnamefont
  {McGuirk}}, \bibinfo {author} {\bibfnamefont {G.~T.}\ \bibnamefont {Foster}},
  \bibinfo {author} {\bibfnamefont {J.~B.}\ \bibnamefont {Fixler}}, \bibinfo
  {author} {\bibfnamefont {M.~J.}\ \bibnamefont {Snadden}}, \ and\ \bibinfo
  {author} {\bibfnamefont {M.~A.}\ \bibnamefont {Kasevich}},\ }\href {\doibase
  10.1103/PhysRevA.65.033608} {\bibfield  {journal} {\bibinfo  {journal} {Phys.
  Rev. A}\ }\textbf {\bibinfo {volume} {65}},\ \bibinfo {pages} {33608}
  (\bibinfo {year} {2002})}\BibitemShut {NoStop}%
\bibitem [{\citenamefont {Sorrentino}\ \emph {et~al.}(2012)\citenamefont
  {Sorrentino}, \citenamefont {Bertoldi}, \citenamefont {Bodart}, \citenamefont
  {Cacciapuoti}, \citenamefont {de~Angelis}, \citenamefont {Lien},
  \citenamefont {Prevedelli}, \citenamefont {Rosi},\ and\ \citenamefont
  {Tino}}]{Sorrentino2012a}%
  \BibitemOpen
  \bibfield  {author} {\bibinfo {author} {\bibfnamefont {F.}~\bibnamefont
  {Sorrentino}}, \bibinfo {author} {\bibfnamefont {A.}~\bibnamefont
  {Bertoldi}}, \bibinfo {author} {\bibfnamefont {Q.}~\bibnamefont {Bodart}},
  \bibinfo {author} {\bibfnamefont {L.}~\bibnamefont {Cacciapuoti}}, \bibinfo
  {author} {\bibfnamefont {M.}~\bibnamefont {de~Angelis}}, \bibinfo {author}
  {\bibfnamefont {Y.-H.}\ \bibnamefont {Lien}}, \bibinfo {author}
  {\bibfnamefont {M.}~\bibnamefont {Prevedelli}}, \bibinfo {author}
  {\bibfnamefont {G.}~\bibnamefont {Rosi}}, \ and\ \bibinfo {author}
  {\bibfnamefont {G.~M.}\ \bibnamefont {Tino}},\ }\href {\doibase
  10.1063/1.4751112} {\bibfield  {journal} {\bibinfo  {journal} {Appl. Phys.
  Lett.}\ }\textbf {\bibinfo {volume} {101}},\ \bibinfo {pages} {114106}
  (\bibinfo {year} {2012})}\BibitemShut {NoStop}%
\bibitem [{\citenamefont {Biedermann}\ \emph {et~al.}(2015)\citenamefont
  {Biedermann}, \citenamefont {Wu}, \citenamefont {Deslauriers}, \citenamefont
  {Roy}, \citenamefont {Mahadeswaraswamy},\ and\ \citenamefont
  {Kasevich}}]{Biedermann2015}%
  \BibitemOpen
  \bibfield  {author} {\bibinfo {author} {\bibfnamefont {G.~W.}\ \bibnamefont
  {Biedermann}}, \bibinfo {author} {\bibfnamefont {X.}~\bibnamefont {Wu}},
  \bibinfo {author} {\bibfnamefont {L.}~\bibnamefont {Deslauriers}}, \bibinfo
  {author} {\bibfnamefont {S.}~\bibnamefont {Roy}}, \bibinfo {author}
  {\bibfnamefont {C.}~\bibnamefont {Mahadeswaraswamy}}, \ and\ \bibinfo
  {author} {\bibfnamefont {M.~A.}\ \bibnamefont {Kasevich}},\ }\href {\doibase
  10.1103/PhysRevA.91.033629} {\bibfield  {journal} {\bibinfo  {journal} {Phys.
  Rev. A}\ }\textbf {\bibinfo {volume} {91}},\ \bibinfo {pages} {33629}
  (\bibinfo {year} {2015})}\BibitemShut {NoStop}%
\bibitem [{\citenamefont {Anandan}(1984)}]{Anandan1984}%
  \BibitemOpen
  \bibfield  {author} {\bibinfo {author} {\bibfnamefont {J.}~\bibnamefont
  {Anandan}},\ }\href@noop {} {\bibfield  {journal} {\bibinfo  {journal} {Phys.
  Rev. A.}\ }\textbf {\bibinfo {volume} {30}},\ \bibinfo {pages} {1615}
  (\bibinfo {year} {1984})}\BibitemShut {NoStop}%
\bibitem [{\citenamefont {Audretsch}\ and\ \citenamefont
  {Marzlin}(1994)}]{Audretsch1994}%
  \BibitemOpen
  \bibfield  {author} {\bibinfo {author} {\bibfnamefont {J.}~\bibnamefont
  {Audretsch}}\ and\ \bibinfo {author} {\bibfnamefont {K.-P.}\ \bibnamefont
  {Marzlin}},\ }\href {\doibase 10.1103/PhysRevA.50.2080} {\bibfield  {journal}
  {\bibinfo  {journal} {Phys. Rev. A}\ }\textbf {\bibinfo {volume} {50}},\
  \bibinfo {pages} {2080} (\bibinfo {year} {1994})}\BibitemShut {NoStop}%
\bibitem [{\citenamefont {Bord\'{e}}\ and\ \citenamefont
  {Lammerzahl}(1996)}]{Borde1996}%
  \BibitemOpen
  \bibfield  {author} {\bibinfo {author} {\bibfnamefont {C.~J.}\ \bibnamefont
  {Bord\'{e}}}\ and\ \bibinfo {author} {\bibfnamefont {C.}~\bibnamefont
  {L\"ammerzahl}},\ }\href@noop {} {\bibfield  {journal} {\bibinfo  {journal}
  {Phys. Bl.}\ }\textbf {\bibinfo {volume} {52}},\ \bibinfo {pages} {238}
  (\bibinfo {year} {1996})}\BibitemShut {NoStop}%
\bibitem [{\citenamefont {Marzlin}\ and\ \citenamefont
  {Audretsch}(1996)}]{Marzlin1996}%
  \BibitemOpen
  \bibfield  {author} {\bibinfo {author} {\bibfnamefont {K.-P.}\ \bibnamefont
  {Marzlin}}\ and\ \bibinfo {author} {\bibfnamefont {J.}~\bibnamefont
  {Audretsch}},\ }\href@noop {} {\bibfield  {journal} {\bibinfo  {journal}
  {Phys. Rev. A}\ }\textbf {\bibinfo {volume} {53}},\ \bibinfo {pages} {312}
  (\bibinfo {year} {1996})}\BibitemShut {NoStop}%
\bibitem [{\citenamefont {Carroll}(2004)}]{Carroll2004a}%
  \BibitemOpen
  \bibfield  {author} {\bibinfo {author} {\bibfnamefont {S.~M.}\ \bibnamefont
  {Carroll}},\ }\href@noop {} {\emph {\bibinfo {title} {{Spacetime and
  Geometry: An Introduction to General Relativity}}}}\ (\bibinfo  {publisher}
  {Addison-Wesley},\ \bibinfo {address} {San Francisco},\ \bibinfo {year}
  {2004})\BibitemShut {NoStop}%
\bibitem [{Note1()}]{Note1}%
  \BibitemOpen
  \bibinfo {note} {Previous work that measured the curvature of the
  gravitational field (rather than potential, as in this work) using three
  separate gravimeters \cite {Rosi2015} was not based on recoil
  effects}\BibitemShut {NoStop}%
\bibitem [{\citenamefont {Dickerson}\ \emph {et~al.}(2013)\citenamefont
  {Dickerson}, \citenamefont {Hogan}, \citenamefont {Sugarbaker}, \citenamefont
  {Johnson},\ and\ \citenamefont {Kasevich}}]{Dickerson2013}%
  \BibitemOpen
  \bibfield  {author} {\bibinfo {author} {\bibfnamefont {S.~M.}\ \bibnamefont
  {Dickerson}}, \bibinfo {author} {\bibfnamefont {J.~M.}\ \bibnamefont
  {Hogan}}, \bibinfo {author} {\bibfnamefont {A.}~\bibnamefont {Sugarbaker}},
  \bibinfo {author} {\bibfnamefont {D.~M.~S.}\ \bibnamefont {Johnson}}, \ and\
  \bibinfo {author} {\bibfnamefont {M.~A.}\ \bibnamefont {Kasevich}},\ }\href
  {\doibase 10.1103/PhysRevLett.111.083001} {\bibfield  {journal} {\bibinfo
  {journal} {Phys. Rev. Lett.}\ }\textbf {\bibinfo {volume} {111}},\ \bibinfo
  {pages} {83001} (\bibinfo {year} {2013})}\BibitemShut {NoStop}%
\bibitem [{\citenamefont {M\"{u}ntinga}\ and\ \citenamefont
  {Others}(2013)}]{Muntinga2013}%
  \BibitemOpen
  \bibfield  {author} {\bibinfo {author} {\bibfnamefont {H.}~\bibnamefont
  {M\"{u}ntinga}}\ \bibinfo {author} {\bibnamefont {et. al.}},\ }\href
  {\doibase 10.1103/PhysRevLett.110.093602} {\bibfield  {journal} {\bibinfo
  {journal} {Phys. Rev. Lett.}\ }\textbf {\bibinfo {volume} {110}},\ \bibinfo
  {pages} {93602} (\bibinfo {year} {2013})}\BibitemShut {NoStop}%
\bibitem [{\citenamefont {McGuirk}\ \emph {et~al.}(2000)\citenamefont
  {McGuirk}, \citenamefont {Snadden},\ and\ \citenamefont
  {Kasevich}}]{McGuirk2000}%
  \BibitemOpen
  \bibfield  {author} {\bibinfo {author} {\bibfnamefont {J.~M.}\ \bibnamefont
  {McGuirk}}, \bibinfo {author} {\bibfnamefont {M.~J.}\ \bibnamefont
  {Snadden}}, \ and\ \bibinfo {author} {\bibfnamefont {M.~A.}\ \bibnamefont
  {Kasevich}},\ }\href {\doibase 10.1103/PhysRevLett.85.4498} {\bibfield
  {journal} {\bibinfo  {journal} {Phys. Rev. Lett.}\ }\textbf {\bibinfo
  {volume} {85}},\ \bibinfo {pages} {4498} (\bibinfo {year}
  {2000})}\BibitemShut {NoStop}%
\bibitem [{\citenamefont {M\"{u}ller}\ \emph {et~al.}(2008)\citenamefont
  {M\"{u}ller}, \citenamefont {Chiow}, \citenamefont {Long}, \citenamefont
  {Herrmann},\ and\ \citenamefont {Chu}}]{Muller2008}%
  \BibitemOpen
  \bibfield  {author} {\bibinfo {author} {\bibfnamefont {H.}~\bibnamefont
  {M\"{u}ller}}, \bibinfo {author} {\bibfnamefont {S.-w.}\ \bibnamefont
  {Chiow}}, \bibinfo {author} {\bibfnamefont {Q.}~\bibnamefont {Long}},
  \bibinfo {author} {\bibfnamefont {S.}~\bibnamefont {Herrmann}}, \ and\
  \bibinfo {author} {\bibfnamefont {S.}~\bibnamefont {Chu}},\ }\href {\doibase
  10.1103/PhysRevLett.100.180405} {\bibfield  {journal} {\bibinfo  {journal}
  {Phys. Rev. Lett.}\ }\textbf {\bibinfo {volume} {100}},\ \bibinfo {pages}
  {180405} (\bibinfo {year} {2008})}\BibitemShut {NoStop}%
\bibitem [{\citenamefont {Chiow}\ \emph {et~al.}(2011)\citenamefont {Chiow},
  \citenamefont {Kovachy}, \citenamefont {Chien},\ and\ \citenamefont
  {Kasevich}}]{Chiow2011}%
  \BibitemOpen
  \bibfield  {author} {\bibinfo {author} {\bibfnamefont {S.-W.}\ \bibnamefont
  {Chiow}}, \bibinfo {author} {\bibfnamefont {T.}~\bibnamefont {Kovachy}},
  \bibinfo {author} {\bibfnamefont {H.-C.}\ \bibnamefont {Chien}}, \ and\
  \bibinfo {author} {\bibfnamefont {M.~A.}\ \bibnamefont {Kasevich}},\ }\href
  {\doibase 10.1103/PhysRevLett.107.130403} {\bibfield  {journal} {\bibinfo
  {journal} {Phys. Rev. Lett.}\ }\textbf {\bibinfo {volume} {107}},\ \bibinfo
  {pages} {130403} (\bibinfo {year} {2011})}\BibitemShut {NoStop}%
\bibitem [{\citenamefont {Kovachy}\ \emph
  {et~al.}(2015{\natexlab{a}})\citenamefont {Kovachy}, \citenamefont
  {Asenbaum}, \citenamefont {Overstreet}, \citenamefont {Donnelly},
  \citenamefont {Dickerson}, \citenamefont {Sugarbaker}, \citenamefont
  {Hogan},\ and\ \citenamefont {Kasevich}}]{Kovachy2015a}%
  \BibitemOpen
  \bibfield  {author} {\bibinfo {author} {\bibfnamefont {T.}~\bibnamefont
  {Kovachy}}, \bibinfo {author} {\bibfnamefont {P.}~\bibnamefont {Asenbaum}},
  \bibinfo {author} {\bibfnamefont {C.}~\bibnamefont {Overstreet}}, \bibinfo
  {author} {\bibfnamefont {C.~A.}\ \bibnamefont {Donnelly}}, \bibinfo {author}
  {\bibfnamefont {S.~M.}\ \bibnamefont {Dickerson}}, \bibinfo {author}
  {\bibfnamefont {A.}~\bibnamefont {Sugarbaker}}, \bibinfo {author}
  {\bibfnamefont {J.~M.}\ \bibnamefont {Hogan}}, \ and\ \bibinfo {author}
  {\bibfnamefont {M.~A.}\ \bibnamefont {Kasevich}},\ }\href {\doibase
  10.1038/nature16155} {\bibfield  {journal} {\bibinfo  {journal} {Nature}\
  }\textbf {\bibinfo {volume} {528}},\ \bibinfo {pages} {530} (\bibinfo {year}
  {2015}{\natexlab{a}})}\BibitemShut {NoStop}%
\bibitem [{\citenamefont {Arndt}\ and\ \citenamefont
  {Hornberger}(2014)}]{Arndt2014}%
  \BibitemOpen
  \bibfield  {author} {\bibinfo {author} {\bibfnamefont {M.}~\bibnamefont
  {Arndt}}\ and\ \bibinfo {author} {\bibfnamefont {K.}~\bibnamefont
  {Hornberger}},\ }\href {\doibase 10.1038/NPHYS2863} {\bibfield  {journal}
  {\bibinfo  {journal} {Nat. Phys.}\ }\textbf {\bibinfo {volume} {10}},\
  \bibinfo {pages} {271} (\bibinfo {year} {2014})}\BibitemShut {NoStop}%
\bibitem [{\citenamefont {Nimmrichter}\ and\ \citenamefont
  {Hornberger}(2013)}]{Nimmrichter2013}%
  \BibitemOpen
  \bibfield  {author} {\bibinfo {author} {\bibfnamefont {S.}~\bibnamefont
  {Nimmrichter}}\ and\ \bibinfo {author} {\bibfnamefont {K.}~\bibnamefont
  {Hornberger}},\ }\href {\doibase 10.1103/PhysRevLett.110.160403} {\bibfield
  {journal} {\bibinfo  {journal} {Phys. Rev. Lett.}\ }\textbf {\bibinfo
  {volume} {110}},\ \bibinfo {pages} {160403} (\bibinfo {year}
  {2013})}\BibitemShut {NoStop}%
\bibitem [{Note2()}]{Note2}%
  \BibitemOpen
  \bibinfo {note} {In the apparatus described here, vibration of the
  retro-reflection mirror used for the atom optics laser beams is the primary
  source of phase noise \cite {Dickerson2013}}\BibitemShut {NoStop}%
\bibitem [{\citenamefont {Gregoire}\ \emph {et~al.}(2015)\citenamefont
  {Gregoire}, \citenamefont {Hromada}, \citenamefont {Holmgren}, \citenamefont
  {Trubko},\ and\ \citenamefont {Cronin}}]{Gregoire2015}%
  \BibitemOpen
  \bibfield  {author} {\bibinfo {author} {\bibfnamefont {M.~D.}\ \bibnamefont
  {Gregoire}}, \bibinfo {author} {\bibfnamefont {I.}~\bibnamefont {Hromada}},
  \bibinfo {author} {\bibfnamefont {W.~F.}\ \bibnamefont {Holmgren}}, \bibinfo
  {author} {\bibfnamefont {R.}~\bibnamefont {Trubko}}, \ and\ \bibinfo {author}
  {\bibfnamefont {A.~D.}\ \bibnamefont {Cronin}},\ }\href {\doibase
  10.1103/PhysRevA.92.052513} {\bibfield  {journal} {\bibinfo  {journal} {Phys.
  Rev. A}\ }\textbf {\bibinfo {volume} {92}},\ \bibinfo {pages} {052513}
  (\bibinfo {year} {2015})}\BibitemShut {NoStop}%
\bibitem [{\citenamefont {Holmgren}\ \emph {et~al.}(2012)\citenamefont
  {Holmgren}, \citenamefont {Trubko}, \citenamefont {Hromada},\ and\
  \citenamefont {Cronin}}]{Holmgren2012}%
  \BibitemOpen
  \bibfield  {author} {\bibinfo {author} {\bibfnamefont {W.~F.}\ \bibnamefont
  {Holmgren}}, \bibinfo {author} {\bibfnamefont {R.}~\bibnamefont {Trubko}},
  \bibinfo {author} {\bibfnamefont {I.}~\bibnamefont {Hromada}}, \ and\
  \bibinfo {author} {\bibfnamefont {A.~D.}\ \bibnamefont {Cronin}},\ }\href
  {\doibase 10.1103/PhysRevLett.109.243004} {\bibfield  {journal} {\bibinfo
  {journal} {Phys. Rev. Lett.}\ }\textbf {\bibinfo {volume} {109}},\ \bibinfo
  {pages} {243004} (\bibinfo {year} {2012})}\BibitemShut {NoStop}%
\bibitem [{\citenamefont {Leonard}\ \emph {et~al.}(2015)\citenamefont
  {Leonard}, \citenamefont {Fallon}, \citenamefont {Sackett},\ and\
  \citenamefont {Safronova}}]{Leonard2015}%
  \BibitemOpen
  \bibfield  {author} {\bibinfo {author} {\bibfnamefont {R.~H.}\ \bibnamefont
  {Leonard}}, \bibinfo {author} {\bibfnamefont {A.~J.}\ \bibnamefont {Fallon}},
  \bibinfo {author} {\bibfnamefont {C.~A.}\ \bibnamefont {Sackett}}, \ and\
  \bibinfo {author} {\bibfnamefont {M.~S.}\ \bibnamefont {Safronova}},\ }\href
  {\doibase 10.1103/PhysRevA.92.052501} {\bibfield  {journal} {\bibinfo
  {journal} {Phys. Rev. A}\ }\textbf {\bibinfo {volume} {92}},\ \bibinfo
  {pages} {052501} (\bibinfo {year} {2015})}\BibitemShut {NoStop}%
\bibitem [{\citenamefont {Arvanitaki}\ \emph {et~al.}(2008)\citenamefont
  {Arvanitaki}, \citenamefont {Dimopoulos}, \citenamefont {Geraci},
  \citenamefont {Hogan},\ and\ \citenamefont {Kasevich}}]{Arvanitaki2008}%
  \BibitemOpen
  \bibfield  {author} {\bibinfo {author} {\bibfnamefont {A.}~\bibnamefont
  {Arvanitaki}}, \bibinfo {author} {\bibfnamefont {S.}~\bibnamefont
  {Dimopoulos}}, \bibinfo {author} {\bibfnamefont {A.~A.}\ \bibnamefont
  {Geraci}}, \bibinfo {author} {\bibfnamefont {J.}~\bibnamefont {Hogan}}, \
  and\ \bibinfo {author} {\bibfnamefont {M.}~\bibnamefont {Kasevich}},\ }\href
  {\doibase 10.1103/PhysRevLett.100.120407} {\bibfield  {journal} {\bibinfo
  {journal} {Phys. Rev. Lett.}\ }\textbf {\bibinfo {volume} {100}},\ \bibinfo
  {pages} {120407} (\bibinfo {year} {2008})}\BibitemShut {NoStop}%
\bibitem [{\citenamefont {Kovachy}\ \emph
  {et~al.}(2015{\natexlab{b}})\citenamefont {Kovachy}, \citenamefont {Hogan},
  \citenamefont {Sugarbaker}, \citenamefont {Dickerson}, \citenamefont
  {Donnelly}, \citenamefont {Overstreet},\ and\ \citenamefont
  {Kasevich}}]{Kovachy2015}%
  \BibitemOpen
  \bibfield  {author} {\bibinfo {author} {\bibfnamefont {T.}~\bibnamefont
  {Kovachy}}, \bibinfo {author} {\bibfnamefont {J.~M.}\ \bibnamefont {Hogan}},
  \bibinfo {author} {\bibfnamefont {A.}~\bibnamefont {Sugarbaker}}, \bibinfo
  {author} {\bibfnamefont {S.~M.}\ \bibnamefont {Dickerson}}, \bibinfo {author}
  {\bibfnamefont {C.~A.}\ \bibnamefont {Donnelly}}, \bibinfo {author}
  {\bibfnamefont {C.}~\bibnamefont {Overstreet}}, \ and\ \bibinfo {author}
  {\bibfnamefont {M.~A.}\ \bibnamefont {Kasevich}},\ }\href {\doibase
  10.1103/PhysRevLett.114.143004} {\bibfield  {journal} {\bibinfo  {journal}
  {Phys. Rev. Lett.}\ }\textbf {\bibinfo {volume} {114}},\ \bibinfo {pages}
  {143004} (\bibinfo {year} {2015}{\natexlab{b}})}\BibitemShut {NoStop}%
\bibitem [{Note3()}]{Note3}%
  \BibitemOpen
  \bibinfo {note} {The finite duration of the optical pulses leads to small
  corrections to calculated distances.}\BibitemShut {Stop}%
\bibitem [{\citenamefont {Sugarbaker}\ \emph {et~al.}(2013)\citenamefont
  {Sugarbaker}, \citenamefont {Dickerson}, \citenamefont {Hogan}, \citenamefont
  {Johnson},\ and\ \citenamefont {Kasevich}}]{Sugarbaker2013}%
  \BibitemOpen
  \bibfield  {author} {\bibinfo {author} {\bibfnamefont {A.}~\bibnamefont
  {Sugarbaker}}, \bibinfo {author} {\bibfnamefont {S.~M.}\ \bibnamefont
  {Dickerson}}, \bibinfo {author} {\bibfnamefont {J.~M.}\ \bibnamefont
  {Hogan}}, \bibinfo {author} {\bibfnamefont {D.~M.~S.}\ \bibnamefont
  {Johnson}}, \ and\ \bibinfo {author} {\bibfnamefont {M.~A.}\ \bibnamefont
  {Kasevich}},\ }\href {\doibase 10.1103/PhysRevLett.111.113002} {\bibfield
  {journal} {\bibinfo  {journal} {Phys. Rev. Lett.}\ }\textbf {\bibinfo
  {volume} {111}},\ \bibinfo {pages} {113002} (\bibinfo {year}
  {2013})}\BibitemShut {NoStop}%
\bibitem [{sup()}]{supplemental}%
  \BibitemOpen
  \href@noop {} {}\bibinfo {note} {See Supplemental Material at [URL
  TBD].}\BibitemShut {Stop}%
\bibitem [{\citenamefont {Dziewonski}\ and\ \citenamefont
  {Anderson}(1981)}]{Dziewonski1981}%
  \BibitemOpen
  \bibfield  {author} {\bibinfo {author} {\bibfnamefont {A.~M.}\ \bibnamefont
  {Dziewonski}}\ and\ \bibinfo {author} {\bibfnamefont {D.~L.}\ \bibnamefont
  {Anderson}},\ }\href@noop {} {\bibfield  {journal} {\bibinfo  {journal}
  {Phys. Earth Planet. Inter.}\ }\textbf {\bibinfo {volume} {25}} (\bibinfo
  {year} {1981})}\BibitemShut {NoStop}%
\bibitem [{\citenamefont {Hogan}\ \emph {et~al.}(2009)\citenamefont {Hogan},
  \citenamefont {Johnson},\ and\ \citenamefont {Kasevich}}]{Hogan2009}%
  \BibitemOpen
  \bibfield  {author} {\bibinfo {author} {\bibfnamefont {J.~M.}\ \bibnamefont
  {Hogan}}, \bibinfo {author} {\bibfnamefont {D.~M.~S.}\ \bibnamefont
  {Johnson}}, \ and\ \bibinfo {author} {\bibfnamefont {M.~A.}\ \bibnamefont
  {Kasevich}},\ }in\ \href {http://arxiv.org/abs/0806.3261} {\emph {\bibinfo
  {booktitle} {Proc. Int. Sch. Phys. "Enrico Fermi" Atom Opt. Sp. Phys.}}},\
  \bibinfo {editor} {edited by\ \bibinfo {editor} {\bibfnamefont
  {E.}~\bibnamefont {Arimondo}}, \bibinfo {editor} {\bibfnamefont
  {W.}~\bibnamefont {Ertmer}}, \ and\ \bibinfo {editor} {\bibfnamefont {W.~P.}\
  \bibnamefont {Schleich}}}\ (\bibinfo  {publisher} {IOS Press},\ \bibinfo
  {address} {Amsterdam},\ \bibinfo {year} {2009})\ pp.\ \bibinfo {pages}
  {411--447},\ \Eprint {http://arxiv.org/abs/0806.3261} {arXiv:0806.3261}
  \BibitemShut {NoStop}%
\bibitem [{\citenamefont {Fixler}\ \emph {et~al.}(2007)\citenamefont {Fixler},
  \citenamefont {Foster}, \citenamefont {Mcguirk},\ and\ \citenamefont
  {Kasevich}}]{Fixler2007a}%
  \BibitemOpen
  \bibfield  {author} {\bibinfo {author} {\bibfnamefont {J.~B.}\ \bibnamefont
  {Fixler}}, \bibinfo {author} {\bibfnamefont {G.~T.}\ \bibnamefont {Foster}},
  \bibinfo {author} {\bibfnamefont {J.~M.}\ \bibnamefont {McGuirk}}, \ and\
  \bibinfo {author} {\bibfnamefont {M.~A.}\ \bibnamefont {Kasevich}},\
  }\href@noop {} {\bibfield  {journal} {\bibinfo  {journal} {Science}\
  }\textbf {\bibinfo {volume} {315}},\ \bibinfo {pages} {74} (\bibinfo {year}
  {2007})}\BibitemShut {NoStop}%
\bibitem [{\citenamefont {Rosi}\ \emph {et~al.}(2014)\citenamefont {Rosi},
  \citenamefont {Sorrentino}, \citenamefont {Cacciapuoti}, \citenamefont
  {Prevedelli},\ and\ \citenamefont {Tino}}]{Rosi2014}%
  \BibitemOpen
  \bibfield  {author} {\bibinfo {author} {\bibfnamefont {G.}~\bibnamefont
  {Rosi}}, \bibinfo {author} {\bibfnamefont {F.}~\bibnamefont {Sorrentino}},
  \bibinfo {author} {\bibfnamefont {L.}~\bibnamefont {Cacciapuoti}}, \bibinfo
  {author} {\bibfnamefont {M.}~\bibnamefont {Prevedelli}}, \ and\ \bibinfo
  {author} {\bibfnamefont {G.~M.}\ \bibnamefont {Tino}},\ }\href@noop {}
  {\bibfield  {journal} {\bibinfo  {journal} {Nature}\ }\textbf {\bibinfo
  {volume} {510}},\ \bibinfo {pages} {518} (\bibinfo {year}
  {2014})}\BibitemShut {NoStop}%
\bibitem [{\citenamefont {Antoine}\ and\ \citenamefont
  {Bord\'{e}}(2003)}]{Antoine2003}%
  \BibitemOpen
  \bibfield  {author} {\bibinfo {author} {\bibfnamefont {C.}~\bibnamefont
  {Antoine}}\ and\ \bibinfo {author} {\bibfnamefont {C.~J.}\ \bibnamefont
  {Bord\'{e}}},\ }\href@noop {} {\bibfield  {journal} {\bibinfo  {journal} {J.
  Opt. B Quantum Semiclassical Opt.}\ }\textbf {\bibinfo {volume} {5}},\
  \bibinfo {pages} {S199} (\bibinfo {year} {2003})}\BibitemShut {NoStop}%
\bibitem [{\citenamefont {Stamper-Kurn}\ \emph {et~al.}(2016)\citenamefont
  {Stamper-Kurn}, \citenamefont {Marti},\ and\ \citenamefont
  {M\"{u}ller}}]{Stamper-Kurn2016}%
  \BibitemOpen
  \bibfield  {author} {\bibinfo {author} {\bibfnamefont {D.~M.}\ \bibnamefont
  {Stamper-Kurn}}, \bibinfo {author} {\bibfnamefont {G.~E.}\ \bibnamefont
  {Marti}}, \ and\ \bibinfo {author} {\bibfnamefont {H.}~\bibnamefont
  {M\"{u}ller}},\ }\href {\doibase 10.1038/nature19108} {\bibfield  {journal}
  {\bibinfo  {journal} {Nature}\ }\textbf {\bibinfo {volume} {537}},\ \bibinfo
  {pages} {E1} (\bibinfo {year} {2016})}\BibitemShut {NoStop}%
\bibitem [{\citenamefont {Kovachy}\ \emph {et~al.}(2016)\citenamefont
  {Kovachy}, \citenamefont {Asenbaum}, \citenamefont {Overstreet},
  \citenamefont {Donnelly}, \citenamefont {Dickerson}, \citenamefont
  {Sugarbaker}, \citenamefont {Hogan},\ and\ \citenamefont
  {Kasevich}}]{Kovachy2016}%
  \BibitemOpen
  \bibfield  {author} {\bibinfo {author} {\bibfnamefont {T.}~\bibnamefont
  {Kovachy}}, \bibinfo {author} {\bibfnamefont {P.}~\bibnamefont {Asenbaum}},
  \bibinfo {author} {\bibfnamefont {C.}~\bibnamefont {Overstreet}}, \bibinfo
  {author} {\bibfnamefont {C.~A.}\ \bibnamefont {Donnelly}}, \bibinfo {author}
  {\bibfnamefont {S.~M.}\ \bibnamefont {Dickerson}}, \bibinfo {author}
  {\bibfnamefont {A.}~\bibnamefont {Sugarbaker}}, \bibinfo {author}
  {\bibfnamefont {J.~M.}\ \bibnamefont {Hogan}}, \ and\ \bibinfo {author}
  {\bibfnamefont {M.~A.}\ \bibnamefont {Kasevich}},\ }\href@noop {} {\bibfield
  {journal} {\bibinfo  {journal} {Nature}\ }\textbf {\bibinfo {volume} {537}},\
  \bibinfo {pages} {E2} (\bibinfo {year} {2016})}\BibitemShut {NoStop}%
\bibitem [{\citenamefont {Hohensee}\ \emph {et~al.}(2012)\citenamefont
  {Hohensee}, \citenamefont {Estey}, \citenamefont {Hamilton}, \citenamefont
  {Zeilinger},\ and\ \citenamefont {M\"{u}ller}}]{Hohensee2012}%
  \BibitemOpen
  \bibfield  {author} {\bibinfo {author} {\bibfnamefont {M.~A.}\ \bibnamefont
  {Hohensee}}, \bibinfo {author} {\bibfnamefont {B.}~\bibnamefont {Estey}},
  \bibinfo {author} {\bibfnamefont {P.}~\bibnamefont {Hamilton}}, \bibinfo
  {author} {\bibfnamefont {A.}~\bibnamefont {Zeilinger}}, \ and\ \bibinfo
  {author} {\bibfnamefont {H.}~\bibnamefont {M\"{u}ller}},\ }\href {\doibase
  10.1103/PhysRevLett.108.230404} {\bibfield  {journal} {\bibinfo  {journal}
  {Phys. Rev. Lett.}\ }\textbf {\bibinfo {volume} {108}},\ \bibinfo {pages}
  {230404} (\bibinfo {year} {2012})}\BibitemShut {NoStop}%
\bibitem [{\citenamefont {Rosi}\ \emph {et~al.}(2015)\citenamefont {Rosi},
  \citenamefont {Cacciapuoti}, \citenamefont {Sorrentino}, \citenamefont
  {Menchetti}, \citenamefont {Prevedelli},\ and\ \citenamefont
  {Tino}}]{Rosi2015}%
  \BibitemOpen
  \bibfield  {author} {\bibinfo {author} {\bibfnamefont {G.}~\bibnamefont
  {Rosi}}, \bibinfo {author} {\bibfnamefont {L.}~\bibnamefont {Cacciapuoti}},
  \bibinfo {author} {\bibfnamefont {F.}~\bibnamefont {Sorrentino}}, \bibinfo
  {author} {\bibfnamefont {M.}~\bibnamefont {Menchetti}}, \bibinfo {author}
  {\bibfnamefont {M.}~\bibnamefont {Prevedelli}}, \ and\ \bibinfo {author}
  {\bibfnamefont {G.~M.}\ \bibnamefont {Tino}},\ }\href {\doibase
  10.1103/PhysRevLett.114.013001} {\ \textbf {\bibinfo {volume} {114}},\
  \bibinfo {pages} {013001} (\bibinfo {year} {2015})}\BibitemShut {NoStop}%
\end{thebibliography}

%

\ifdefined\WORDCOUNT
    \end{document}
\else
\fi

\end{document}